\documentclass{Interspeech2024}

\usepackage{cite}
\usepackage{amsfonts}
\usepackage{algorithmic}
\usepackage{multirow}
\usepackage{makecell}
\usepackage{xcolor}
\usepackage{subcaption}
\interspeechcameraready

\title{Training Universal Vocoders with Feature Smoothing-\polish{B}ased Augmentation Methods for High-\polish{Q}uality TTS Systems}
\name[affiliation={1}]{Jeongmin}{Liu}
\name[affiliation={1}]{Eunwoo}{Song}

\address{$^1$NAVER Cloud Corp., Seongnam, Korea}
\email{jeongmin.liu@navercorp.com, eunwoo.song@navercorp.com}


\newcommand{\jedittwo}[1]{\textcolor{black}{#1}}
\newcommand{\polish}[1]{\textcolor{black}{#1}}
\newcommand{\eedit}[1]{\textcolor{black}{#1}}
\newcommand{\jedit}[1]{\textcolor{black}{#1}}

\begin{document}
%\fontsize{9}{11}\selectfont
\fontsize{8.7}{10.6}\selectfont
%\fontsize{8.6}{10.5}\selectfont

\maketitle

% the abstract here must exactly match the abstract entered into the paper submission system
\begin{abstract}
    While universal vocoders have achieved proficient waveform generation across diverse voices, their integration into text-to-speech (TTS) tasks often results in degraded synthetic quality.
    To address this challenge, we present a novel augmentation technique for training universal vocoders.
    Our training scheme randomly applies linear smoothing filters to input acoustic features, facilitating vocoder generalization across a wide range of smoothings.
    It significantly mitigates the training-inference mismatch, enhancing the naturalness of synthetic output even when the acoustic model produces overly smoothed features.
    Notably, our method is applicable to any vocoder without requiring architectural modifications or dependencies on specific acoustic models. 
    The experimental results validate the superiority of our vocoder over conventional methods, achieving 11.99\% and 12.05\% improvements in mean opinion scores when integrated with Tacotron 2 and FastSpeech 2 TTS acoustic models, respectively.

\end{abstract}

\section{Introduction}

    Recent advancements in modeling capacity have led to the development of universal vocoders capable of generating high-fidelity speech waveforms.
    Trained on diverse audio samples recorded from various environments, these vocoding models accommodate a wide spectrum of voices, languages, and styles \cite{jang2021univnet,lee2023bigvgan,song2023dspgan}. 
    However, integrating them with text-to-speech (TTS) acoustic models remains challenging due to separate training processes, potentially resulting in degraded synthesis quality caused by the acoustic model's tendency to produce overly smoothed features.

    One straightforward solution is to fine-tune the vocoder using acoustic features generated by the corresponding acoustic model \cite{shen2018tacotron2}.
    However, this approach compromises the vocoder's universality, \polish{because} different acoustic models trained by different speakers or styles necessitate distinct fine-tuned vocoders, requiring substantial deployment resources and time.
    Alternatively, fully end-to-end TTS models have been proposed \cite{kim2021vits,lim2022jets,tan2024naturalspeech}, \polish{whereby the} joint training of the acoustic and vocoding models can avoid the training-inference mismatch problem.
    However, this approach may lose the flexibility to control the acoustic characteristics of \if 0 (polisih) the \fi output speech, as the model often relies on implicit latent representations instead of explicit acoustic features.

    To address these limitations, we propose a novel feature augmentation strategy to preserve the vocoder's universality while mitigating quality loss within the TTS framework.
    This method involves applying linear smoothing filters to input acoustic features, \eedit{approximating} their distributions with those generated by the acoustic model (i.e., smoothed along the time and frequency axes). 
    Specifically, the size of \polish{the} smoothing filters is randomly selected for every training step, exposing the target vocoder to various levels of smoothing.
    Consequently, the vocoder gains enhanced adaptability to the acoustic model's over-smoothing problem without necessitating fine-tuning or architectural modifications.

    Our proposed method offers the advantage of applicability to any type of universal vocoder.
    In particular, we focus on the UnivNet model
    \if 0
    \footnote
    {
        In the experiments, the effectiveness of the proposed method is also verified with the HiFi-GAN vocoder \cite{kong2020hifi} as well.
    } 
    \fi
    \cite{jang2021univnet}, a generative adversarial network (GAN)-based universal neural vocoder. 
    We enhance its performance with the additional harmonic-noise (HN) architecture \cite{hwang2021mbhnpwg,xu2022refinegan,huang2022singgan} for stable harmonic production and two discriminators, namely\polish{,} multi-scale short-time Fourier transform (MS-STFT) \cite{defossez2023encodec} and collaborative multi-band (CoMB) \cite{bak2023avocodo}\polish{,} for improving training accuracy.
    The experimental results demonstrate that our universal vocoder trained with the proposed method outperforms the traditional approaches across different TTS systems.

    \begin{figure}[t]
        % \vspace{-0.015\linewidth}
        \centering
        \begin{subfigure}[b]{0.475\linewidth}
            \centering
            \includegraphics[width=0.99\linewidth]{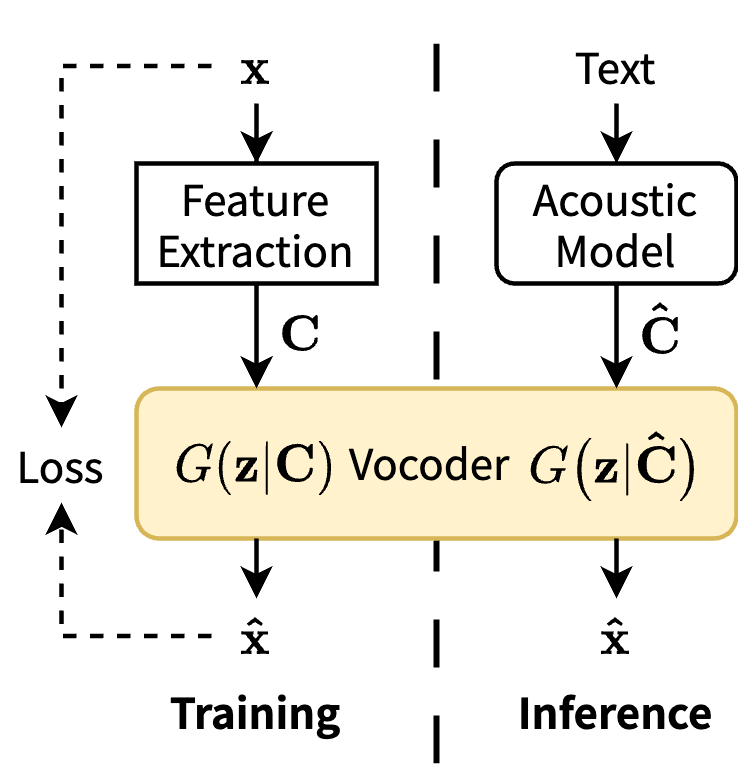}
            \caption{}
            \label{fig:method:conventional}
        \end{subfigure}
        \hspace{0.005\linewidth}
        \begin{subfigure}[b]{0.494\linewidth}
            \centering
            \includegraphics[width=0.99\linewidth]{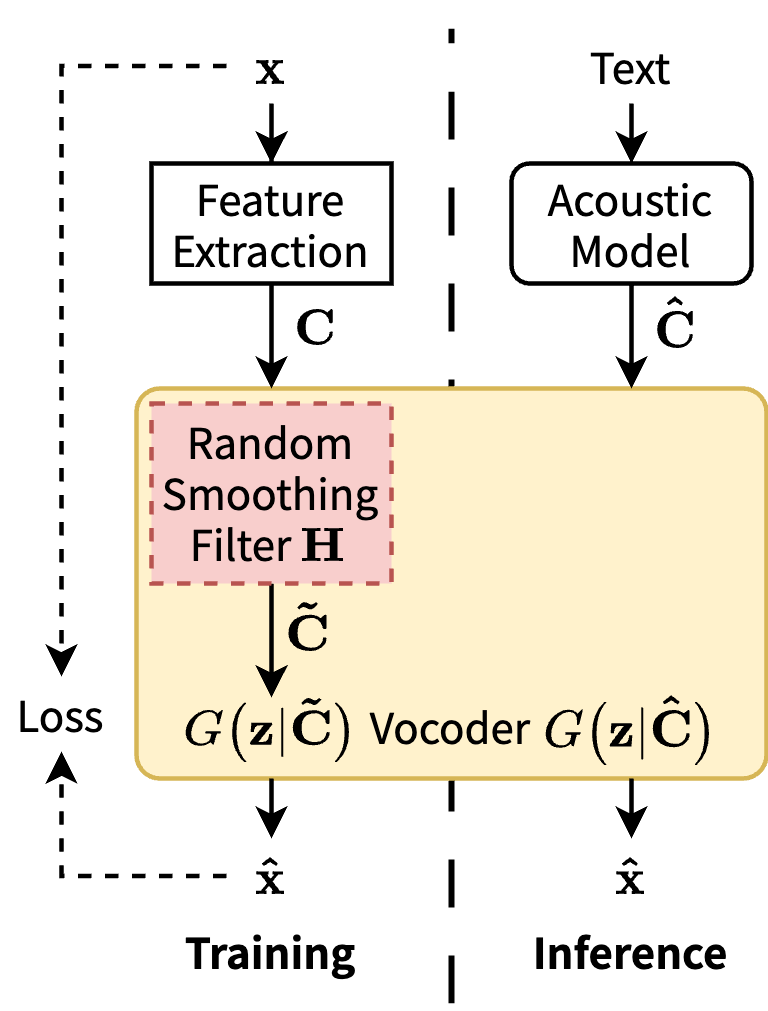}
            \caption{}
            \label{fig:method:proposed}
        \end{subfigure}
        % \vspace{-2mm}
        \caption{
            Block diagram of the vocoding process in the TTS framework: (a) \if 0 (polish) The \fi conventional and (b) \if 0 (polish) the \fi proposed methods.
        }
        % \vspace{-3mm}
    \end{figure}
    
\section{Method\polish{s}}
\label{sec:method}
\subsection{Neural vocoder\polish{s}}
\label{ssec:vocoder}

    GAN-based neural vocoders \cite{kumar2019melgan,yamamoto2020parallel} are generative models that \polish{comprise} a generator $G$ and a discriminator $D$ \cite{goodfellow2014gan}.
    \polish{As i}llustrated in Figure~\ref{fig:method:conventional}, these models are designed to learn the distribution of time-domain speech waveforms $\mathbf x$ conditioned on their corresponding ground-truth acoustic features $\mathbf C$\polish{,} with the following objectives: 
    \begin{align}
        \label{eq:objectiveG}
        &\min_G \mathbb{E}_{\mathbf z \sim \mathcal{N}(\mathbf 0, \mathbf I), (\mathbf x,\mathbf C)}
            \left[ L_{G}(\mathbf z, \mathbf x, \mathbf C) \right], \\
        \label{eq:objectiveD}
        &\min_D \mathbb{E}_{\mathbf z \sim \mathcal{N}(\mathbf 0, \mathbf I), (\mathbf x,\mathbf C)}
            \left[ L_{D}(\mathbf z, \mathbf x, \mathbf C) \right],
    \end{align}
    where $L_G$ and $L_D$ are the following losses of the generator and the discriminator, respectively:
    \begin{align}\label{eq:loss}
        L_{G}(\mathbf z, \mathbf x, \mathbf C)
            &= d\left[ D(G(\mathbf z | \mathbf C)), 1 \right] + \lambda L_{\mathrm{aux}}(G(\mathbf z | \mathbf C), \mathbf x), \\
        L_{D}(\mathbf z, \mathbf x, \mathbf C)
            &= d\left[ D(\mathbf x), 1 \right] + d\left[ D(G(\mathbf z |\mathbf C)), 0 \right],
    \end{align}
    where $d$ is a distance function (e.g., $L_2$ distance), and $\lambda$ is the weight for an auxiliary loss $L_{\mathrm{aux}}$.

    \begin{figure}[t]
        \centering
        \includegraphics[width=0.95\linewidth]{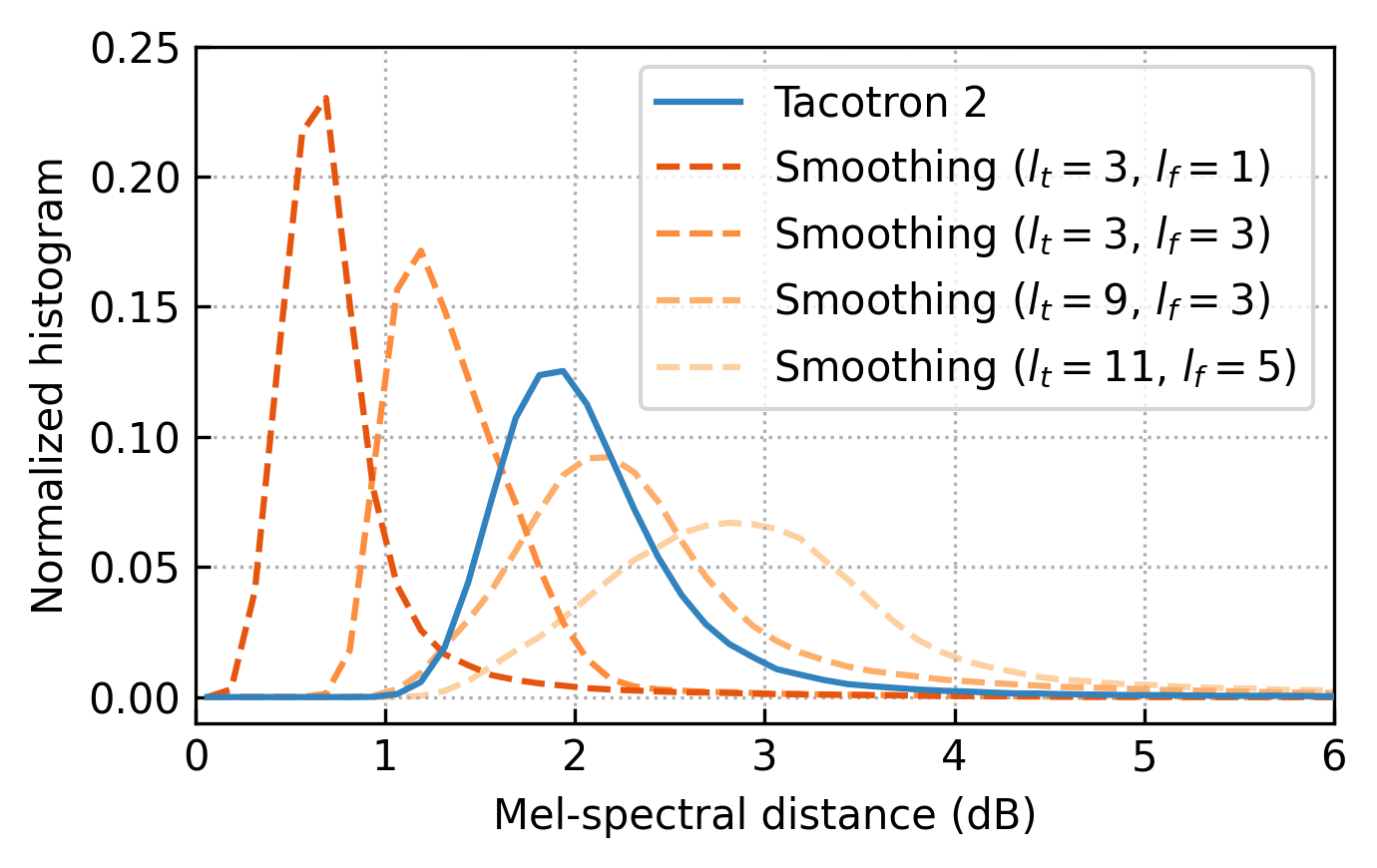}        
    
        \vspace{-2mm}
        \caption{
            Normalized histograms of the \polish{mel-spectral distance (MSD)} between the ground-truth mel-spectrograms and those generated by Tacotron 2, or simulated using different sizes of smoothing filters.
            These results were obtained from the test set.
        }
        \label{fig:smoothing_histogram}
        \vspace{-3mm}
    \end{figure}

\subsection{Proposed smoothing augmentation method}
\label{ssec:smoothing}

    Previous research has demonstrated the technical feasibility of universal vocoding through the utilization of numerous training samples \cite{jang2021univnet,lee2023bigvgan,song2023dspgan}. However, \if 0 (polish) the \fi TTS systems often struggle to produce high-quality speech due to the exposure bias problem.
    This issue arises because the vocoder, having been trained solely on ground-truth mel-spectrograms, lacks robustness in handling the over-smoothing tendencies of acoustic models.
    
    To address this challenge, we propose a smoothing augmentation method designed to mitigate the exposure bias problem \polish{effectively} within the TTS framework. 
    As shown in Figure~\ref{fig:method:proposed}, the key idea is to train the universal vocoder conditioned on smoothed mel-spectrograms that closely mimic the distribution of those generated by the acoustic models.
    To approximate the target distribution, we simulate the conditional acoustic feature by applying a linear smoothing filter\polish{,} as follows:
    \begin{equation}\label{eq:filtering}
        \mathbf{\tilde C}=\mathbf{H}*\mathbf{C},
    \end{equation}
    where $*$ denotes the 2-dimensional convolution operation and $\mathbf{H}$ represents the smoothing filter designed to cover a range of smoothings from the acoustic models.
    \eedit{Instead of using \if 0 (polish) the \fi ground-truth mel-spectrograms, the universal vocoder is conditioned on these smoothed features and optimized with modified objectives\polish{,} as follows:}
    %Instead of using the ground-truth mel-spectrograms, these smoothed features are conditioned \jcomment{feature가 conditioned되는 게 아니라 conditioning하는 것 같습니다} to train the universal vocoder, optimizing \jcomment{optimizing에 대한 주어동사가 these smoothed features are 인가요?} the model with modified objectives as follows:
    \begin{align}
        \label{eq:objectiveG_prop}
        &\min_G \mathbb{E}_{\mathbf z, (\mathbf x, \mathbf{\tilde C})}
            \left[ L_{G}(\mathbf z, \mathbf x, \mathbf{\tilde C}) \right], \\
        \label{eq:objectiveD_prop}
        &\min_D \mathbb{E}_{\mathbf z, (\mathbf x, \mathbf{\tilde C})}
            \left[ L_{D}(\mathbf z, \mathbf x, \mathbf{\tilde C}) \right].
    \end{align}
    This process \polish{does not} require any fine-tuning procedure or modification of the network architecture, allowing \if 0 (polish) for \fi compact training without additional deployment resources.
    Furthermore, as our training scheme exposes the vocoder to a diverse range of smoothings, the model becomes more generalized to the acoustic model's smoothing errors in the generation step\polish{. T}hus, the entire TTS system is empowered to produce more natural speech outputs.

    \begin{figure}[t]
        \centering
        \begin{subfigure}[t]{0.45\linewidth}
            \centering
            \includegraphics[height=29.5mm]{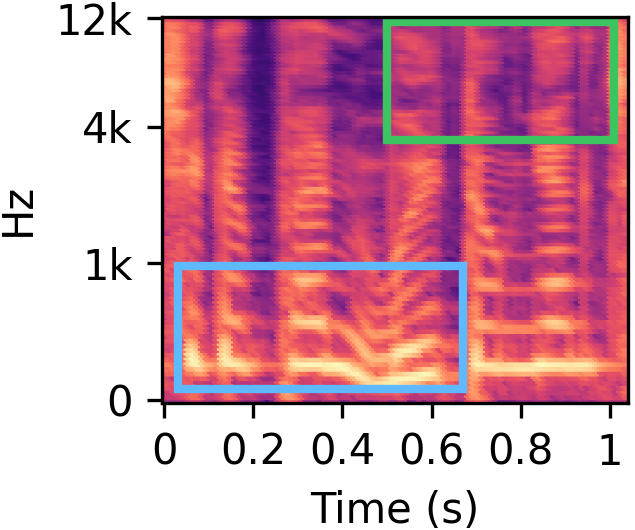}
            \captionsetup{margin={23pt,0pt}}
            \caption{}
            %\caption{The ground-truth $\mathbf C$}
        \end{subfigure}
        \begin{subfigure}[t]{0.54\linewidth}
            \centering
            \includegraphics[height=29.5mm]{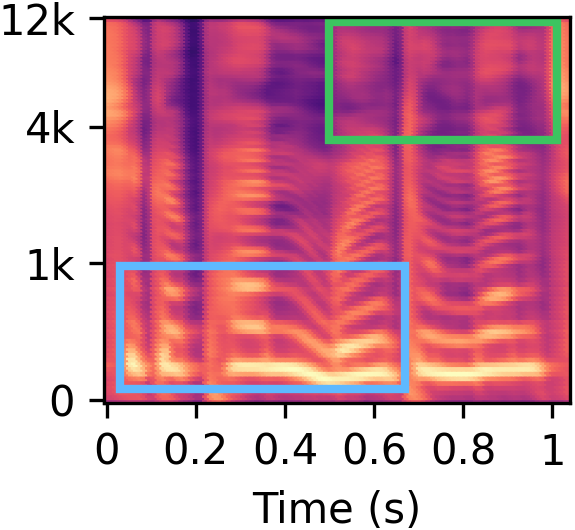}
            \includegraphics[height=29.5mm]{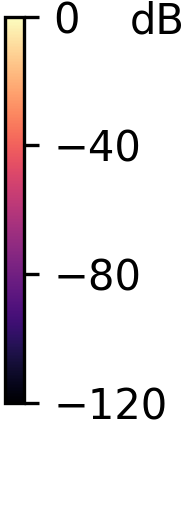}
            \captionsetup{margin={0pt,15pt}}
            \caption{}
            %\caption{$\mathbf{\hat C}$, predicted by Tacotron2}
        \end{subfigure} 

        \begin{subfigure}[t]{0.45\linewidth}
            \centering
            \includegraphics[height=29.5mm]{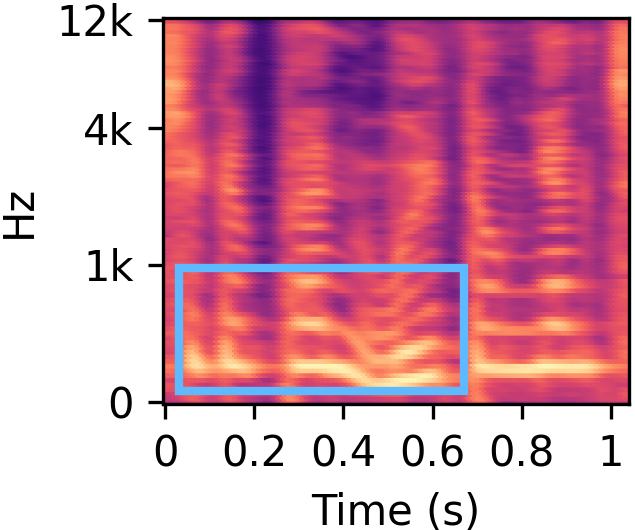}
            \captionsetup{margin={23pt,0pt}}
            \caption{}
            %\caption{
            %    $\mathbf{\tilde C}$, smoothed by the filter of \\
            %    $(l_t,l_f)=(5,1)$
            %}
        \end{subfigure}
        \begin{subfigure}[t]{0.54\linewidth}
            \centering
            \includegraphics[height=29.5mm]{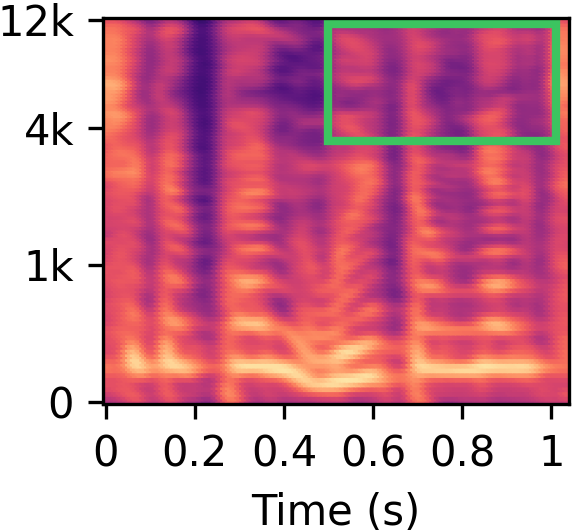}
            \includegraphics[height=29.5mm]{fig/colorbar_6_4.png}
            \captionsetup{margin={0pt,15pt}}
            \caption{}
            %\caption{
            %    $\mathbf{\tilde C}$, smoothed by the filter of \\
            %    $(l_t,l_f)=(5,3)$
            %}
        \end{subfigure}
        \vspace{-2mm}
        \caption{
            Mel-spectrograms of (a) \if 0 (polish) the \fi ground-truth speech, (b) generated by the Tacotron 2 acoustic model, (c) simulated using the smoothing filter ($l_t$=5, $l_f$=1), and (d) simulated using another filter ($l_t$=5, $l_f$=3). The rectangular areas highlight instances \polish{in which} our method simulates smoothings similar to those that occurred by the acoustic model.
        }
        \label{fig:smoothing_example}
        \vspace{-3mm}
    \end{figure}

    \begin{figure*}
        \centering

        \begin{subfigure}[b]{0.39\textwidth}
            \centering
            \includegraphics[width=0.99\linewidth]{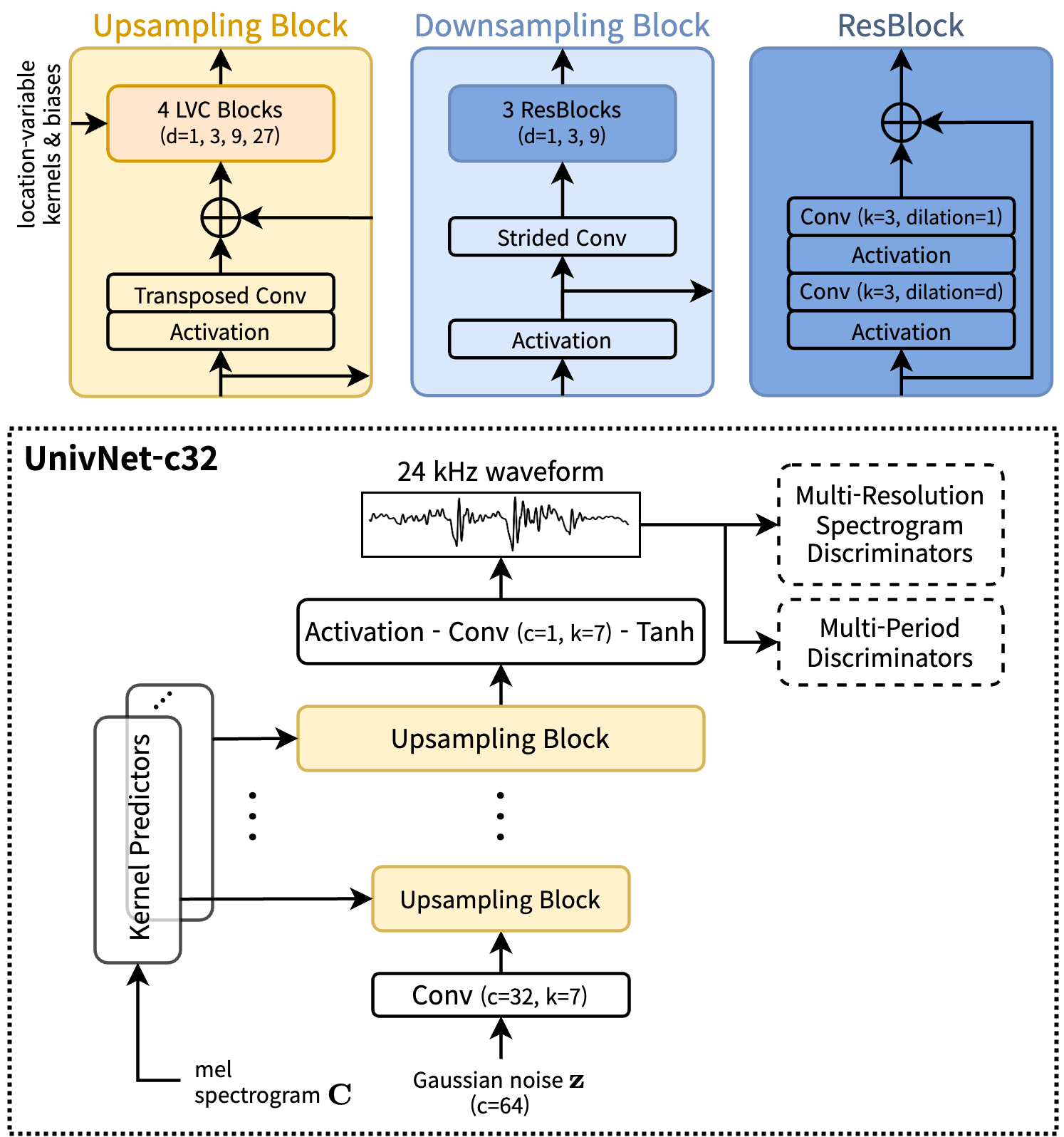}
            \vspace{-4.5mm}
            \caption{}
            \label{fig:network:univnet}
        \end{subfigure}
        \hspace{1mm}
        \begin{subfigure}[b]{0.40\textwidth}
            \centering
            \includegraphics[width=0.99\linewidth]{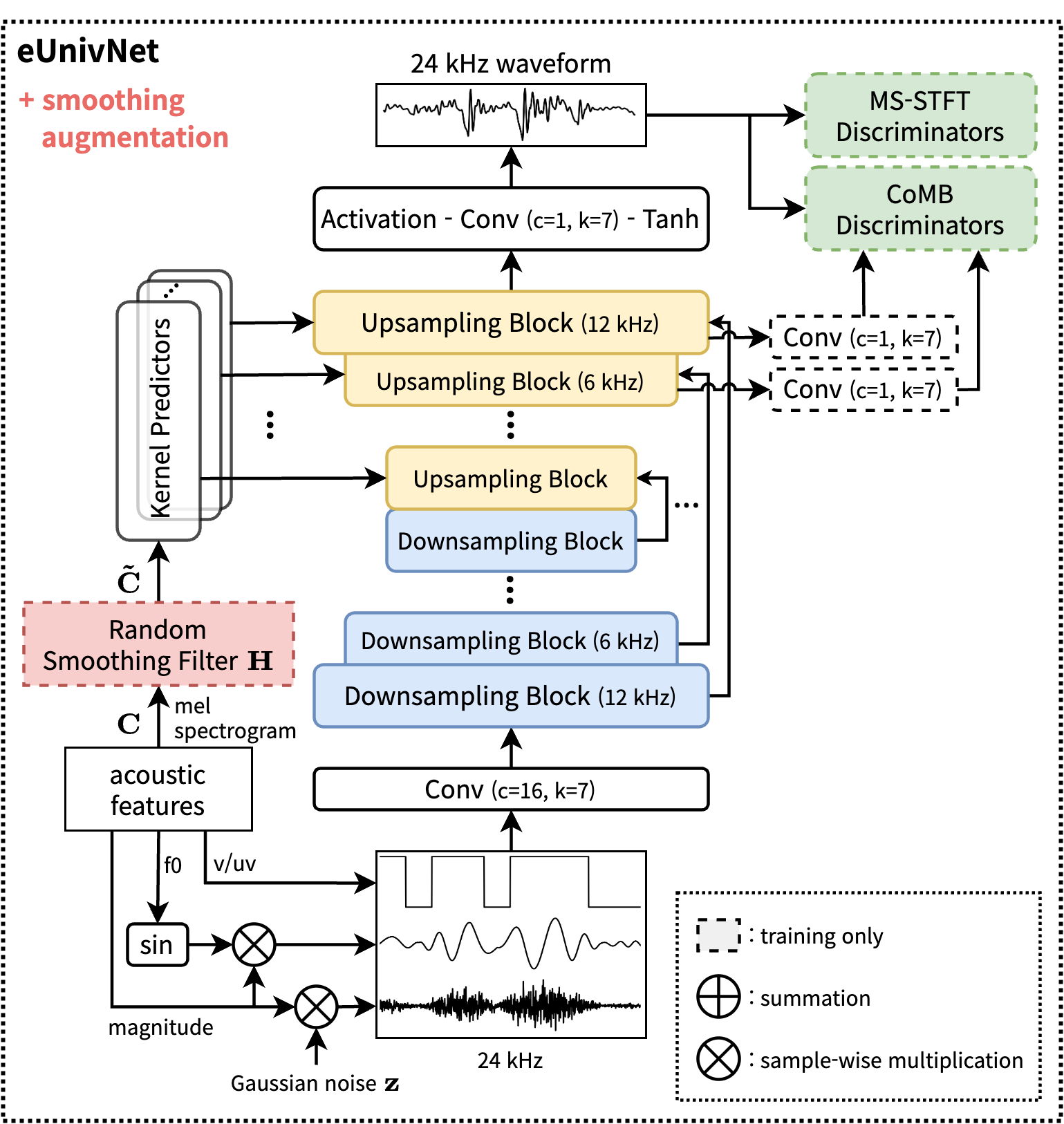}
            \vspace{-4.5mm}
            \caption{}
            \label{fig:network:harnovo}
        \end{subfigure}
        \vspace{-2mm}
        \caption{
            The UnivNet architectures: (a) \polish{t}he vanilla UnivNet-c32 model and (b) the proposed eUnivNet model. 
            The notations $c$ and $k$ denote the number of channels and the kernel size of the convolution layer, respectively.
            %The notations $c$, $k$, and $d$ denote the number of channels, the kernel size of the convolution layer, and the dilation factor of the dilated convolution layer, respectively.
            %The neural network architecture of (a) the original UnivNet-c32 and (b) our improved version.
            %$c$ and $k$ are the number of channels and the kernel size of the convolution layer, respectively, and $d$ is the dilation factor of the dilated convolution.
            %The activation functions in the original and the improved version are LeakyReLU and Snake, respectively.
        }
        \label{fig:network}
        \vspace{-3.5mm}
    \end{figure*}
        
\subsubsection{Filter design}
\label{sssec:filter}

    Among various types of smoothing filters, i.e., $\mathbf H$ in \polish{E}quation~(\ref{eq:filtering}), we employ a 2-dimensional triangular\footnote
    {
        In our preliminary experiments, it was noticeable that any type of linear LPF, e.g., rectangular LPF, could be used to compose $\mathbf H$.
    }
    low-pass filter (LPF), defined as follows:
    \begin{equation} 
        \label{eq:filter}
        h_{t,f}
            =\frac{\left\lceil l_t/2 \right\rceil - \big|\, t - \left\lceil l_t/2 \right\rceil \big|}{ \left\lceil l_t/2 \right\rceil^2 } \cdot \frac{\left\lceil l_f/2 \right\rceil - \big|\, f - \left\lceil l_f/2 \right\rceil \big|}{ \left\lceil l_f/2 \right\rceil^2 },
    \end{equation}
    where $l_t$ and $l_f$ denote the filter sizes along \if 0 (polish) the \fi time frame $t$ and frequency bin $f$, respectively.

    To enhance the vocoder's robustness across a diverse range of smoothings, it is crucial to randomly vary the filter sizes $l_t$ and $l_f$ for every training step.
    Specifically, these parameters are randomly sampled based on the following distributions:
    \begin{equation}\label{eq:filter_size}
        l_t \sim p(l; N_t),\ l_f \sim p(l; N_f),
    \end{equation}   
    where $N_t$ and $N_f$ denote the numbers of possible candidates for $l_t$ and $l_f$, respectively;
    $p(l; N)$ denotes \polish{a distribution that is mostly uniform, except for} the \eedit{non-smoothing case} ($l$=1)\polish{,} as follows: 
    \begin{align}\label{eq:filter_size2}
        &p(l; N) =
        \begin{cases}
            p_g & \text{for }\ l=1, \\
            p_s & \text{for }\ l \in \{ 3, 5, \cdots, 2N-1 \},
        \end{cases}\\
        &\text{where }\ p_g+(N-1)p_s = 1.
    \end{align}
    While $p_g$ and $p_s$ can have the same value, our early experiments indicated that increasing $p_g$ beyond $p_s$, such as $p_g$=$2/3$, yielded improvements in the vocoder's synthetic quality.
    %Notably, we set the filter sizes $l_t$ and $l_f$ to odd numbers to ensure symmetric filters and use six and three different filter sizes for $N_t$ and $N_f$, respectively, in the subsequent experiments outlined in the following sections.
    Notably, we set the filter sizes $l_t$ and $l_f$ to odd numbers to ensure symmetric filters.
    %; \eedit{for instance, we used $l_t$=11 and $l_f$=5} in the subsequent experiments outlined in the following sections.

\subsubsection{Analysis of smoothing augmentation}

    To validate the generalization capacity of our proposed method, we examine\polish{d} the mel-spectral distance (MSD; dB)\polish{,} defined as the L2-norm of mel-spectral frames between the ground-truth mel-spectrogram and those predicted by the acoustic model or simulated by our smoothing filters.    
    In Figure~\ref{fig:smoothing_histogram}, the \textit{orange dashed} lines depict examples of MSD histograms corresponding to \polish{distinct} sizes of smoothing filters. 
    \polish{The r}andom sampling \polish{of} these filter sizes during training \eedit{guides} the vocoder to accommodate various levels of smoothing.
    We believe that this enables the vocoder to effectively manage the overly smoothed features generated by acoustic models such as Tacotron 2 \cite{shen2018tacotron2}, as indicated by the \textit{blue solid} line.
    This observation is further supported by the mel-spectrograms presented in Figure~\ref{fig:smoothing_example}, where the simulated features exhibit \polish{tendencies similar} to those generated by the Tacotron 2 acoustic model.
    \if 0 (polish) in both low- and high-frequency regions 는 삭제함 \fi

    \begin{table*}[t]
        \begin{center}
        \caption{
            TTS naturalness MOS results with 95\% confidence intervals with respect to different training strategies (\textit{seen} cases): \jedittwo{\if 0 (polish) The \fi conventional separate training (ST)}, fine-tuning (FT) by generated features\jedittwo{,} and the proposed smoothing augmentation (SA) method.
            % The best MOS scores are in bold.
        }
        \vspace{-3mm}
        \label{table:main}
        {\footnotesize %\small
        \begin{tabular}{ccccccccccc}
        \Xhline{2\arrayrulewidth}
        \multirow{2}{*}{System} & Acoustic & \multirow{2}{*}{Vocoder} & \multirow{2}{*}{ST} & \multirow{2}{*}{FT} & \multirow{2}{*}{SA} & \multicolumn{5}{c}{MOS ($\uparrow$)} \\ 
                                & Model    & & & & & F1 & F2 & M1 & M2 & Average      \\
        \hline
        \hline
        S1 & T2 & eUnivNet       & \checkmark & - & - & 3.93$\pm$0.11 & 3.54$\pm$0.11 & 3.59$\pm$0.12 & 3.64$\pm$0.12 & 3.67$\pm$0.06 \\	
        S2 & T2 & eUnivNet       & - & \checkmark & - & 4.48$\pm$0.09 & 4.10$\pm$0.11 & 4.26$\pm$0.11 & 4.10$\pm$0.12 & 4.23$\pm$0.05 \\	
        S3 & T2 & eUnivNet       & \checkmark & - & \checkmark & 4.30$\pm$0.10 & 3.93$\pm$0.11 & 4.07$\pm$0.11 & 4.13$\pm$0.11  & 4.11$\pm$0.05 \\	
        \hline
        S4 & T2 & eUnivNet-HN-G  & \checkmark & - & \checkmark & 3.77$\pm$0.12 & 3.31$\pm$0.13 & 3.43$\pm$0.13 & 3.69$\pm$0.12 & 3.55$\pm$0.06 \\	
        S5 & T2 & eUnivNet-M/C-D & \checkmark & - & \checkmark & 3.90$\pm$0.11 & 3.69$\pm$0.11 & 3.73$\pm$0.11 & 3.48$\pm$0.12 & 3.70$\pm$0.06 \\	
        S6 & T2 & UnivNet-c32    & \checkmark & - & \checkmark & 3.72$\pm$0.11 & 3.57$\pm$0.12 & 3.53$\pm$0.13 & 3.62$\pm$0.12 & 3.61$\pm$0.06 \\
        \hline
        S7 & FS2 & eUnivNet      & \checkmark & - & - & 3.44$\pm$0.13 & 2.86$\pm$0.12 & 2.93$\pm$0.14 & 3.06$\pm$0.12 & 3.07$\pm$0.06 \\	
        S8 & FS2 & eUnivNet      & \checkmark & - & \checkmark & 3.84$\pm$0.12 & 3.14$\pm$0.11 & 3.28$\pm$0.12 & 3.51$\pm$0.12  & 3.44$\pm$0.06 \\	
        \hline
        S9 & T2 & HiFi-GAN V1    & \checkmark & - & - & 2.54$\pm$0.13 & 2.24$\pm$0.13 & 1.98$\pm$0.12 & 2.23$\pm$0.12 & 2.25$\pm$0.06 \\	
        S10 & T2 & HiFi-GAN V1   & \checkmark & - & \checkmark & 3.87$\pm$0.10 & 3.66$\pm$0.11 & 3.60$\pm$0.12 & 3.55$\pm$0.11 & 3.67$\pm$0.06 \\	
        \hline
        Recording & - & -     & -   & - & -       & 4.48$\pm$0.10 & 4.37$\pm$0.11 & 4.27$\pm$0.12 & 4.23$\pm$0.10 & 4.34$\pm$0.06 \\
        \Xhline{2\arrayrulewidth}
        \end{tabular}}	
        \end{center}
        \vspace{-5mm}
    \end{table*}
    
\section{Experiments}
\subsection{Datasets}
    The universal vocoder was trained on an internal dataset comprising recordings in five languages (Korean, Japanese, \polish{Mandarin Chinese}, English, and Spanish) spoken by 73 speakers.
    The dataset contained 219,407 utterances (about 269 hours), with 5\% reserved for validation.
    Each waveform was sampled at 24 kHz and quantized by 16 bits.    
    For acoustic model training, we randomly selected four Korean speakers (two female, F1 and F2, and two male, M1 and M2) from the vocoder's training dataset (seen speakers) and two Korean speakers (one female, F3, and one male, M3) not included in the training set (unseen speakers).
    The corpus for seen speakers comprised 4,600 utterances (about 7.7 hours), while the corpus for unseen speakers contained 2,400 utterances (about 3.6 hours).
    Validation and testing utilized 6\% and 3\% of each corpus, respectively.

    We extracted 100-dimensional mel-spectrograms, covering \if 0 (polish) from \fi 0 to 12 kHz, with a 256-length frame shift and a 1,024-length Hann window.
    Additionally, log-F0 and voicing flags, extracted using the PYIN algorithm \cite{mauch2014pyin}, were included to compose 102-dimensional acoustic features.
    \polish{Before} training, all acoustic features were globally normalized using the mean and variance of the training set.
    
    \begin{table}[t]
        \begin{center}
        \caption{
            \hspace{-0.17em}Vocoding model details\polish{,} including model size and inference speed: The abbreviations HN-G and M/C-D represent HN-generator and MS-STFT/CoMB discriminators, respectively. The inference speed indicates \jedittwo{how many times} faster \polish{a model can generate waveforms} than real-time. The evaluation was conducted on a single core of Intel Xeon Gold 5120 2.20 GHz.
            % The best MOS scores are in bold.
        }
        \vspace{-3mm}
        \label{table:models}
        {\footnotesize %\small
        \begin{tabular}{ccccc}
        \Xhline{2\arrayrulewidth}
        \multirow{2}{*}{Vocoder} & \multirow{2}{*}{HN-G} & \multirow{2}{*}{M/C-D} & Model             & Inference \\ 
                                 &                       &                        & Size ($\downarrow$) & Speed ($\uparrow$) \\
        \hline
        \hline
        eUnivNet (ours)          & \checkmark            & \checkmark             & 5.32 M            & $\times$2.54    \\ 
        eUnivNet-HN-G            & \checkmark            & -                      & 5.32 M            & $\times$2.54    \\ 
        eUnivNet-M/C-D           & -                     & \checkmark             & 5.30 M            & $\times$2.83    \\ 
        UnivNet-c32 \cite{jang2021univnet} & -           & -                      & 14.79 M           & $\times$1.47    \\ 
        HiFi-GAN V1 \cite{kong2020hifi}    & -           & -                      & 14.00 M           & $\times$0.48    \\ 
        \Xhline{2\arrayrulewidth}
        \end{tabular}}	
        \end{center}
        \vspace{-6mm}
    \end{table}

\subsection{Model details}
\label{ssec:modelsetups}
\subsubsection{Vocoding model}

    Despite the availability of many state-of-the-art vocoders, we opted for the UnivNet model \cite{jang2021univnet}\polish{,} thanks to its competitive synthetic quality and fast generation speed.
    Our model followed the overall setup of the original UnivNet-c16 model\footnote{
        We used an open\polish{-}source implementation at the following URL:\\ \url{https://github.com/maum-ai/univnet/}\polish{.}
    }, but we enhanced it by incorporating the following techniques, as depicted in Figure 4b:
    \polish{First}, we introduced a \textit{harmonic-noise} (HN) model into the generator \cite{hwang2021mbhnpwg,xu2022refinegan,huang2022singgan}.
    The model received three inputs \eedit{composed of} F0-dependent sinusoidal, Gaussian noise, and a sequence of voicing information to enable the generator to efficiently learn the periodic and aperiodic behavior of the target waveform.
    We employed U-Net-style downsampling blocks \cite{ronneberger2015unet} to align the \eedit{sample-level} inputs with \polish{the} \eedit{frame-level} mel-spectrogram.
    %As these inputs were sampled at every 24 kHz \jcomment{at every 24kHz가 일반적인 표현인가요?}, we employed U-Net-style downsampling blocks \cite{ronneberger2015unet} to align them with conditional mel-spectrograms.
    \polish{Second}, the discriminators were replaced with MS-STFT and CoMB discriminators, extending the vocoder's capabilities to capture complex audio features.
    The MS-STFT discriminator \cite{defossez2023encodec} facilitated analysis in the complex STFT domain, while the CoMB discriminator \cite{bak2023avocodo} allowed \if 0 (polish) for \fi \polish{the} capturing \polish{of the} frequency band-wise periodic attributes of the target voice.
    These modifications were integrated into the model \if 0 (polish) , \fi following the\polish{ir} official implementations \cite{defossez2023encodec,bak2023avocodo}. 
    Additionally, adjustments were made to the upsampling ratios of \jedit{the upsampling} blocks from \{8,8,4\} to \{8,8,2,2\} \if 0 (polish), \fi to connect the last two blocks to the CoMB discriminator.
    \polish{Last}, all activation functions in the generator were replaced with the Snake function \cite{larochelle2020snake}, known for its effectiveness in handling periodic signals \cite{lee2023bigvgan}.
    We define\polish{d} our enhanced vocoder as an \textit{eUnivNet} for the remaining part\polish{s} of the experiments.

    The generator and discriminators were trained for 600k steps using the AdamW optimizer \cite{loshchilov2018adamw}.
    During training, \eedit{the model was conditioned on} the ground-truth \jedit{acoustic features} for the first 450k steps \eedit{and} \jedit{the features with randomly smoothed mel-spectrograms} derived from our proposed method for the remaining 150k steps.  % melspec만 smooth되는 것 언급 필요
    \eedit{When sampling the size of smoothing filters in \polish{E}quation~(\ref{eq:filter_size}), we used six and three candidates for $N_t$ and $N_f$, respectively.}
    Additional training parameters included a weight decay of 0.01, a batch size of 32, and an exponentially decayed learning rate from $10^{-4}$ with a decay rate of 0.99 per epoch.
    %(approximately 6,300 steps).

\subsubsection{Acoustic model}
    To generate acoustic \jedit{features} from the text, we employed Tacotron 2 (T2) with a phoneme-alignment approach \cite{okamoto2019tacotron} due to its stable generation and competitive synthetic quality.        
    The model received 364-dimensional phoneme-level linguistic features as inputs \eedit{and predicted the corresponding phoneme duration} through a combination of \eedit{three} fully connected \eedit{layers} and \eedit{one} long short-term memory (LSTM) \eedit{network}.
    By utilizing this predicted duration, \eedit{the linguistic features} were upsampled to the frame \if 0 (polish) - \fi level and \eedit{transformed into high-level context features via three convolutional layers\polish{,} followed by a bi-directional LSTM network}.
    \eedit{Consequently}, the T2 decoder autoregressively decoded \eedit{those context features} to reconstruct the target acoustic features.
    %The model received 364-dimensional phoneme-level linguistic features as inputs, which were transformed into high-level context features through a combination of fully connected (FC) and long short-term memory (LSTM) layers. \jcomment{확인해보니 364 ling feat $\rightarrow$ Length regulator (duration from DM) $\rightarrow$ 3Conv+1BiLSTM(the same as original T2) $\rightarrow$ original T2 Decoder 입니다}
    %Simultaneously, the same linguistic features were fed into a duration model, comprising three FC layers and one LSTM layer, to predict the corresponding phoneme duration.
    %By utilizing this predicted duration, the high-level context features were upsampled to the frame-level and autoregressively decoded by the T2 decoder to reconstruct the target acoustic features. 
    The model was initialized by Xavier initializer \cite{xavier2010init} and trained by Adam optimizer \cite{diederik2014adam}.
    %More detailed setups were given in the conventional work \cite{song2020neural}.
    % in our previous work

    To faithfully evaluate the generalization capacity of the proposed method, we included a FastSpeech 2 (FS2) acoustic model \cite{ren2020fastspeech2} as \polish{the} baseline.
    The FS2 model \eedit{was} trained similarly to the T2 model, \eedit{but it} used non-\polish{autoregressive} encoder and decoder.
    More detailed setups \eedit{for training \polish{the} T2 and FS2 models} were given in \if 0 (polish) the \fi conventional work \cite{hwang2021tts}.

\subsection{Evaluations}
    % MOS setup (2월 첫주에는 맡겨보자)
    % wiki page: https://wiki.navercorp.com/pages/viewpage.action?pageId=1603960422

    We performed naturalness mean opinion score (MOS) tests to evaluate the TTS quality of the proposed method.    
    \if 0
        For each speaker, we randomly selected ten utterances from the test set and synthesized speech samples via TTS systems with different universal vocoders as follows:
        %
        \begin{itemize}
        \item {\bf eUnivNet}: Proposed enhanced UnivNet with both the HN-generator and the MS-STFT/Co-MB discriminators
        \item {\bf eUnivNet-G}: eUnivNet only with the HN-generator
        \item {\bf eUnivNet-D}: eUnivNet only with the MS-STFT/Co-MB discriminators
        %\item {\bf UnivNet}: Baseline vanilla UnivNet-c32\footnote{https://github.com/maum-ai/univnet} model \cite{jang2021univnet}
        \item {\bf UnivNet}: Baseline vanilla UnivNet-c32\footnote{
            \eedit{We used an open source implementation at the following URL: \\} \url{https://github.com/maum-ai/univnet/}
        } model \cite{jang2021univnet}
        %\item {\bf HiFi-GAN}: Baseline HiFi-GAN V1\footnote{https://github.com/jik876/hifi-gan} model \cite{kong2020hifi}
        \item {\bf HiFi-GAN}: Baseline HiFi-GAN V1 model \cite{kong2020hifi}.
        \end{itemize}
        %
        Twenty native Korean listeners were asked to rate the synthetic quality (in total, 10 utterances $\times$ 4 speakers $\times$ 20 listeners = 800 hits for each system) using the \polish{5}-point responses: \polish{1=Bad, 2=Poor, 3=Fair, 4=Good, and 5=Excellent}.
    \fi
    Twenty native Korean listeners were asked to rate the synthetic quality using \if 0 (polish) the \fi \polish{5}-point responses: \polish{1=Bad, 2=Poor, 3=Fair, 4=Good, and 5=Excellent}.
    For each speaker, we randomly selected \polish{10} utterances\footnote{
        Generated audio samples are available at the following URL:\\ \url{https://sytronik.github.io/demos/voc_smth_aug}\polish{.}
    } from the test set (in total, \polish{10 utterances$\times$4 speakers$\times$20 listeners=800 hits} for each system).
    The speech samples were synthesized by the different vocoders\polish{,} as described in Table~\ref{table:models}.
    
    Table~\ref{table:main} shows the evaluation results with respect to \polish{various} systems\polish{,} and the analy\polish{tic} results are summarized as follows:
    Compared to the vanilla UnivNet-c32 (S6), our eUnivNet model (S3) performed significantly better despite having half the number of convolutional channels (e.g., 32 vs. 16).
    This highlights the importance of employing the HN-generator (S4) and the MS-STFT/CoMB discriminators (S5) to improve \if 0 (polish) the \fi synthetic quality.
    \polish{Noteworthily,} although the overall MOS score of the eUnivNet-HN-G model was lower than that of the vanilla UnivNet-c32, it offered benefits in terms of lower model size (36.0\%) and fast inference speed (172.8\%)\polish{,} as described in Table~\ref{table:models}.

    %Among the eUnivNet-based vocoders (S1, S2, and S3), \jedittwo{the conventional universal vocoder} training strategy (S1) performed the worst due to the exposure bias problem. 
    Among the eUnivNet-based vocoders (S1, S2, and S3), \jedittwo{the conventional} training method (S1), \jedittwo{i.e., separated from the acoustic model,} performed the worst due to the exposure bias problem.
    Fine-tuning the vocoder with the generated \eedit{features} (S2) significantly addressed this limitation\polish{,} but required time-consuming deployment resources\polish{,} such as generating all \jedit{features} in the training set and \jedit{retraining the vocoder speaker-dependently}.
    \polish{Conversely}, a single model trained solely with our smoothing augmentation method (S3) provided competitive synthetic quality without any dependency related to the speaker or acoustic model.
    
    \polish{The g}eneralized performance was verified by changing the acoustic model (S7 vs. S8) and the vocoder (S9 vs. S10)\polish{. T}he proposed method significantly improved the naturalness of synthesized speech.
    This trend was also observed in \polish{the} other experiments in Table~\ref{table:unseen}, where the proposed method robustly generated unseen speakers' voices compared to conventional methods.

    \begin{table}[t]
        \begin{center}
        \caption{
            TTS naturalness MOS results with 95\% confidence intervals (\textit{unseen} cases).
            % The best MOS scores are in bold.
        }
        \vspace{-3mm}
        \label{table:unseen}
        {\footnotesize %\small      
        \begin{tabular}{cccc}
        \Xhline{2\arrayrulewidth}
        \multirow{2}{*}{System} & \multicolumn{3}{c}{MOS ($\uparrow$)} \\ 
                                & F3 & M3 & Average\\
        \hline
        \hline
        S1 & 3.10$\pm$0.13 & 3.49$\pm$0.12 & 3.29$\pm$0.09 \\
        S3 & 3.67$\pm$0.13 & 3.86$\pm$0.11 & 3.76$\pm$0.08 \\
        \hline
        Recording  & 3.99$\pm$0.12 & 4.39$\pm$0.11 & 4.19$\pm$0.08 \\
        \Xhline{2\arrayrulewidth}
        \end{tabular}}	
        \end{center}
    \vspace{-6mm}
    \end{table}
    
\section{Conclusion}
    This paper propose\polish{s} a novel feature smoothing augmentation method for training universal vocoders aimed at mitigating the mismatch between the acoustic model and \polish{the} vocoder within the TTS framework.
    Our method introduced \if 0 (polish) the \fi random linear filters to augment acoustic features, thereby approximating their distributions to those generated by acoustic models. 
    This approach enhances the generalization capacity of universal vocoders, enabling the generation of high-quality speech outputs even when the acoustic model produces overly smoothed features.
    \polish{The e}xperimental results verified the superiority of our vocoder over \if 0 (polish) the \fi conventional methods.
    Future research directions should explore extending this framework to other generation tasks\polish{,} such as singing voice synthesis and music/audio generation.

\newpage

% will be added after acceptance
\section{Acknowledgements}
We would like to thank Min-Jae Hwang, Meta AI, Seattle, WA, USA, for the helpful discussion.
This work was supported by Voice, NAVER Cloud Corp., Seongnam, Korea.

\bibliographystyle{IEEEtran}
\bibliography{mybib}

% Generated by IEEEtran.bst, version: 1.13 (2008/09/30)
\begin{thebibliography}{10}
\providecommand{\url}[1]{#1}
\csname url@samestyle\endcsname
\providecommand{\newblock}{\relax}
\providecommand{\bibinfo}[2]{#2}
\providecommand{\BIBentrySTDinterwordspacing}{\spaceskip=0pt\relax}
\providecommand{\BIBentryALTinterwordstretchfactor}{4}
\providecommand{\BIBentryALTinterwordspacing}{\spaceskip=\fontdimen2\font plus
\BIBentryALTinterwordstretchfactor\fontdimen3\font minus \fontdimen4\font\relax}
\providecommand{\BIBforeignlanguage}[2]{{%
\expandafter\ifx\csname l@#1\endcsname\relax
\typeout{** WARNING: IEEEtran.bst: No hyphenation pattern has been}%
\typeout{** loaded for the language `#1'. Using the pattern for}%
\typeout{** the default language instead.}%
\else
\language=\csname l@#1\endcsname
\fi
#2}}
\providecommand{\BIBdecl}{\relax}
\BIBdecl

\bibitem{jang2021univnet}
W.~Jang, D.~Lim, J.~Yoon, B.~Kim, and J.~Kim, ``{UnivNet}: A neural vocoder with multi-resolution spectrogram discriminators for high-fidelity waveform generation,'' in \emph{Proc. INTERSPEECH}, 2021, pp. 2207--2211.

\bibitem{lee2023bigvgan}
S.~gil Lee, W.~Ping, B.~Ginsburg, B.~Catanzaro, and S.~Yoon, ``Big{VGAN}: A universal neural vocoder with large-scale training,'' in \emph{Proc. ICLR}, 2023.

\bibitem{song2023dspgan}
K.~Song, Y.~Zhang, Y.~Lei, J.~Cong, H.~Li, L.~Xie, G.~He, and J.~Bai, ``{DSPGAN}: A {GAN}-based universal vocoder for high-fidelity {TTS} by time-frequency domain supervision from {DSP},'' in \emph{Proc. ICASSP}, 2023, pp. 1--5.

\bibitem{shen2018tacotron2}
J.~Shen, R.~Pang, R.~J. Weiss, M.~Schuster, N.~Jaitly, Z.~Yang, Z.~Chen, Y.~Zhang, Y.~Wang, R.~Skerrv-Ryan, R.~A. Saurous, Y.~Agiomvrgiannakis, and Y.~Wu, ``Natural {TTS} synthesis by conditioning {WaveNet} on mel spectrogram predictions,'' in \emph{Proc. ICASSP}, 2018, pp. 4779--4783.

\bibitem{kim2021vits}
J.~Kim, J.~Kong, and J.~Son, ``Conditional variational autoencoder with adversarial learning for end-to-end text-to-speech,'' in \emph{Proc. ICML}, 2021, pp. 5530--5540.

\bibitem{lim2022jets}
D.~Lim, S.~Jung, and E.~Kim, ``{JETS}: Jointly training {FastSpeech2} and {HiFi-GAN} for end to end text to speech,'' in \emph{Proc. INTERSPEECH}, 2022, pp. 21--25.

\bibitem{tan2024naturalspeech}
X.~Tan, J.~Chen, H.~Liu, J.~Cong, C.~Zhang, Y.~Liu, X.~Wang, Y.~Leng, Y.~Yi, L.~He, S.~Zhao, T.~Qin, F.~Soong, and T.-Y. Liu, ``{NaturalSpeech}: End-to-end text-to-speech synthesis with human-level quality,'' \emph{IEEE Trans. Pattern Analysis and Machine Intelligence}, pp. 1--12, 2024.

\bibitem{hwang2021mbhnpwg}
M.-J. Hwang, R.~Yamamoto, E.~Song, and J.-M. Kim, ``High-fidelity {Parallel WaveGAN} with multi-band harmonic-plus-noise model,'' in \emph{Proc. INTERSPEECH}, 2021, pp. 2227--2231.

\bibitem{xu2022refinegan}
S.~Xu, W.~Zhao, and J.~Guo, ``{RefineGAN}: Universally generating waveform better than ground truth with highly accurate pitch and intensity responses,'' in \emph{Proc. INTERSPEECH}, 2022, pp. 1591--1595.

\bibitem{huang2022singgan}
R.~Huang, C.~Cui, F.~Chen, Y.~Ren, J.~Liu, Z.~Zhao, B.~Huai, and Z.~Wang, ``{SingGAN}: Generative adversarial network for high-fidelity singing voice generation,'' in \emph{Proc. ACM Multimedia}, 2022, pp. 2525--2535.

\bibitem{defossez2023encodec}
A.~D{\'e}fossez, J.~Copet, G.~Synnaeve, and Y.~Adi, ``High fidelity neural audio compression,'' \emph{Transactions on Machine Learning Research}, 2023.

\bibitem{bak2023avocodo}
T.~Bak, J.~Lee, H.~Bae, J.~Yang, J.-S. Bae, and Y.-S. Joo, ``Avocodo: Generative adversarial network for artifact-free vocoder,'' in \emph{Proc. AAAI}, vol.~37, no.~11, Jun. 2023, pp. 12\,562--12\,570.

\bibitem{kumar2019melgan}
K.~Kumar, R.~Kumar, T.~de~Boissiere, L.~Gestin, W.~Z. Teoh, J.~Sotelo, A.~de~Br{\'e}bisson, Y.~Bengio, and A.~C. Courville, ``{MelGAN}: {G}enerative adversarial networks for conditional waveform synthesis,'' in \emph{Proc. NeurIPS}, 2019, pp. 14\,881--14\,892.

\bibitem{yamamoto2020parallel}
R.~Yamamoto, E.~Song, and J.~Kim, ``Parallel {W}ave{GAN}: A fast waveform generation model based on generative adversarial networks with multi-resolution spectrogram,'' in \emph{Proc. ICASSP}, 2020, pp. 6199--6203.

\bibitem{goodfellow2014gan}
I.~Goodfellow, J.~Pouget-Abadie, M.~Mirza, B.~Xu, D.~Warde-Farley, S.~Ozair, A.~Courville, and Y.~Bengio, ``Generative adversarial nets,'' in \emph{Proc. NeurIPS}, 2014.

\bibitem{mauch2014pyin}
M.~Mauch and S.~Dixon, ``{PYIN}: A fundamental frequency estimator using probabilistic threshold distributions,'' in \emph{Proc. ICASSP}, 2014, pp. 659--663.

\bibitem{kong2020hifi}
J.~Kong, J.~Kim, and J.~Bae, ``{HiFi-GAN}: {G}enerative adversarial networks for efficient and high fidelity speech synthesis,'' in \emph{Proc. NeurIPS}, 2020, pp. 17\,022--17\,033.

\bibitem{ronneberger2015unet}
O.~Ronneberger, P.~Fischer, and T.~Brox, ``{U-Net}: Convolutional networks for biomedical image segmentation,'' in \emph{Proc. MICCAI}, 2015, pp. 234--241.

\bibitem{larochelle2020snake}
L.~Ziyin, T.~Hartwig, and M.~Ueda, ``Neural networks fail to learn periodic functions and how to fix it,'' in \emph{Proc. NeurIPS}, 2020, pp. 1583--1594.

\bibitem{loshchilov2018adamw}
I.~Loshchilov and F.~Hutter, ``Decoupled weight decay regularization,'' in \emph{Proc. ICLR}, 2019.

\bibitem{okamoto2019tacotron}
T.~Okamoto, T.~Toda, Y.~Shiga, and H.~Kawai, ``Tacotron-based acoustic model using phoneme alignment for practical neural text-to-speech systems,'' in \emph{Proc. ASRU}, 2019, pp. 214--221.

\bibitem{xavier2010init}
X.~Glorot and Y.~Bengio, ``Understanding the difficulty of training deep feedforward neural networks,'' in \emph{Proc. AISTATS}, 2010, pp. 249--256.

\bibitem{diederik2014adam}
\BIBentryALTinterwordspacing
D.~P. Kingma and J.~Ba, ``Adam: {A} method for stochastic optimization,'' \emph{CoRR}, vol. abs/1412.6980, 2014. [Online]. Available: \url{http://arxiv.org/abs/1412.6980}
\BIBentrySTDinterwordspacing

\bibitem{ren2020fastspeech2}
Y.~Ren, C.~Hu, T.~Qin, S.~Zhao, Z.~Zhao, and T.-Y. Liu, ``{FastSpeech} 2: {F}ast and high-quality end-to-end text-to-speech,'' in \emph{Proc. ICLR}, 2021.

\bibitem{hwang2021tts}
M.-J. Hwang, R.~Yamamoto, E.~Song, and J.-M. Kim, ``{TTS-by-TTS}: {TTS}-driven data augmentation for fast and high-quality speech synthesis,'' in \emph{Proc. ICASSP}, 2021, pp. 6598--6602.

\end{thebibliography}

\end{document}